# Topological flatband loop states in fractal-like photonic lattices

Limin Song[1,6], Yuqing Xie[1,6], Shiqi Xia[1], Liqin Tang[1,2,4], Daohong Song[1,2,5], Jun-Won Rhim[3], and Zhigang Chen[1,2]

[1] *The MOE Key Laboratory of Weak-Light Nonlinear Photonics, TEDA Applied Physics Institute and School of Physics, Nankai University, Tianjin 300457, China*
[2] *Collaborative Innovation Center of Extreme Optics, Shanxi University, Taiyuan, Shanxi 030006, People's Republic of China*
[3] *Department of Physics, Ajou University, Suwon, 16499, Korea*
[4] *tanya@nankai.edu.cn,* [5] *songdaohong@nankai.edu.cn,*
[6] *These authors contributed equally to this work*
*Corresponding author: zgchen@nankai.edu.cn*

**Abstract:**

Noncontractible loop states (NLSs) are recently realized topological entity in flatband lattices, arising typically from band touching at a point where a flat band intersects one or more dispersive bands. There exists also band touching across a plane, where one flat band overlaps another all over the Brillouin zone without crossing a dispersive band. Such isolated plane-touching flat bands remain largely unexplored. For example, what are the topological features associated with such flatband degeneracy? Here, we demonstrate for the first time to our knowledge nontrivial NLSs and robust boundary modes in a system with such degeneracy. Based on a tailored photonic lattice constructed from the well-known fractal Sierpinski gasket, we theoretically analyze the wavefunction singularities and the conditions for the existence of the NLSs. We show that the NLSs can exist in both singular and nonsingular flat bands, as a direct reflection of the real-space topology. Experimentally, we observe directly such flatband NLSs in a laser-written Corbino-shaped fractal-like lattice. This work not only leads to a deep understanding of the mechanism behind the nontrivial flatband states, but also opens up new avenues to explore fundamental phenomena arising from the interplay of flatband degeneracy, fractal structures and band topology.

## 1. Introduction

Flat bands are energy bands characterized by the vanishing dispersion and immobile charge carriers, hosting an ideal stage to explore strongly-correlated phenomena [1, 2]. The study of flatband model has a venerable history in condensed matter physics [3, 4], but has recently attracted a great deal of interest, particularly in photonics [5-19], exciton-polaritons [20-22], sythentic electronic structures [23, 24], and Bose-Einstein condenses [25, 26]. Indeed, in such flatband systems, a variety of fundamental phenomena have been demonstrated, ranging from Landau-Zener Bloch oscillations, superfluidity, superconductivity in twisted bilayer graphene, to quantum distance and anomalous Landau levels [27-32]. Equally intriguing are the flatband systems that lack periodicity or Hermiticity, such as moiré lattices with "extremely" flat higher bands, aperiodic and hyperbolic flatband lattices, and flat bands in non-Hermitian systems [33-39]. Recently, the exploration of flatband localization has been extended to synthetic space of frequency dimensions [40] and in Creutz superradiance lattices with tunable synthetic gauge fields [41]. Thus far, those studies have mainly explored systems with the so-called "*point-touching*" of the Bloch bands, in which one flat band touches dispersive band(s) at a quadratic or linearly conical intersection. Exemplary examples include Kagome [8, 14, 24, 42, 43], Lieb [5-7, 10, 44], and super-honeycomb [13, 45, 46] lattices. The flat bands with such degeneracy are often considered topologically trivial in momentum space due to the canted or trivial winding of the flatband pseudospin structure [30]. The corresponding Bloch wave function exhibits singularities (discontinuities) with unquantized Berry phase [47]. Even if the value of the Berry phase can be quantized due to certain crystalline symmetry, the band flatness enforces it to be zero [30, 47-49]. In many flatband systems, however, one can encounter flatband degeneracy with a "*line-touching*" where type-II and type-III Dirac cones are involved [50], or even a "*plane-touching*" where one flat band overlaps another flat band all over the Brillouin zone [18, 19, 35, 42, 48]. Such systems bring about a host of interesting questions in the context of flatband physics. For example, what are the new features of wave localization from degenerate flat bands, and can plane-touching degeneracy support nontrivial flatband states mediated by real-space topology? Here we explore and address some of these questions.

It is instructive to recall some basic consensuses about flatband physics first, especially with respect to flatband localized states. As is well known, under a tight-binding model with a finite hopping range, a flatband system can support well-confined, localized states, often referred to as the *compact localized states* (CLSs) [2, 11, 51, 52], as demonstrated extensively in photonic lattices [5-8, 15-19] and polariton-exciton condensates [21, 22]. All flatband states, which are compactly localized within a finite spatial region but having exactly zero amplitudes outside of it, can be called the CLSs. But in general, it is often chosen a relatively small size as a basic CLS, which can be linearly independent of other basic CLSs with the same shape but different positions, to construct a wave-function set spanning the flat band. From the perspective of state counting, a flatband lattice consisting of $N$ unit cells, in principle, should support $N$ linearly independent CLSs [42]. However, if there exists a singular band touching [48] with other dispersive bands, one can only find $(N-1)$ linearly independent CLSs [14, 42]. Therefore, some non-compact (or compact only in one dimension) eigenstates should be included in addition to the CLSs to complete the flatband basis. Such non-compact modes were found to be the so-called *noncontractible loop states* (NLSs) winding around the entire (infinite) lattices [14, 42, 48], or unconventional flatband line states in the finite lattice system with tailored boundary conditions [10, 13]. The NLSs cannot be obtained by linear superposition of the conventional CLSs, and they cannot be continuously deformed into the CLSs in a torus geometry [14, 48]. Furthermore, there exist two such NLSs in a two-dimensional (2D) system, so that there are in fact $(N+1)$ degenerate eigenstates. This usually means that the flat band should touch other dispersive band(s) at certain points in momentum space. It was theoretically and experimentally found that the NLSs and unconventional flatband line states are inherent to the point-touching degeneracy of the singular flat band, protected by real-space topology and featured by an *immovable discontinuity* in the corresponding Bloch wave functions [13, 14, 42, 48, 49]. In addition to the unconventional flatband line states [10, 13], robust boundary modes (RBMs), which also exist in open boundary geometries, are thought to have the same real-space topological origin as the NLSs according to a novel bulk-boundary correspondence [14, 48, 49].

Recently, we have realized fractal-like photonic lattices by using the continuous-wave (CW) laser-writing technique [10, 18]. Such fractal-like lattices were previously conceived from a fractal geometry - the well-known Sierpinski gasket [35, 53], exhibiting simultaneously point (singular) and plane (nonsingular) band touching. In such photonic lattices, we have observed distinct but basic types of CLSs for both singular and nonsingular flat bands [18]. In a similar vein, research efforts have been made to realize other forms of the CLSs by using the femtosecond laser-writing technique [19]. However, nontrivial flatband localized states, especially the NLSs that manifest the real-space topology, have not been investigated thus far in the fractal-like lattices, either in theory or experiment. Intrinsically, a plane-touching degeneracy is a combination of an infinite number of point-touching degeneracies, thus it is natural to ask whether flatband plane-touching can give rise to NLSs and other nontrivial flatband states, and whether these localized states represent the characteristics of real-space topology?

In this work, we demonstrate nontrivial flatband states in the 2D fractal-like photonic lattices, tailored with different open boundaries or into a Corbino-shaped geometry with periodic boundary conditions. We focus on a set of examples of the localized states (NLSs and RBMs) in such fractal-like photonic lattices, and in particular those supported by spectrally isolated, plane-touching flat bands.

These nontrivial flatband states are fundamentally different from conventional CLSs previously observed [18, 19]. Our theoretical analysis shows that, in fractal-like photonic lattices with a plane-touching degeneracy, the existence of the NLSs depends on the selection of the basic CLSs to span the whole flat band. For those NLSs that do exist, we observe directly in a laser-written fractal-like lattice in Corbino geometry, and corroborate our experimental observations by numerical simulation. While our work is based on a specific photonic system, the scheme and concepts developed here are appliable to other flatband systems beyond photonics.

## 2. Main results

*Complete and incomplete spanning sets on the compact localized states*

We start by considering a fractal-like photonic lattice comprising an array of periodically arranged evanescently coupled waveguides as shown Fig. 1(a), and a corresponding experimentally established lattice laser-written in a photorefractive crystal [10, 18] is displayed in Fig. 1(b). There are six sublattices (the first generation of the Sierpinski gasket [35], labelled by A~F ) per unit cell as marked in Fig. 1(a), which naturally give six energy bands. Figure 1(c) depicts the calculated band structure $\beta(k_x, k_y)$ based on the tight-binding model. One can find that the top dispersive bands and the bottom flat band are gapped. From the solution of the band structure (see **Supplementary Materials (SM)** for details), we know that there exists a zero-energy flat band $\beta_3 = 0$, which touches its neighboring linearly dispersive bands $\beta_2$ and $\beta_4$ at a point, forming a pseudospin-1 Dirac-like cone similar to the Lieb [10, 54, 55] and super-honeycomb [13, 45] lattices. Differently, the present lattice shows no sublattice symmetry. The bottom isolated flat band $\beta_{5,6} = -2$ is doubly-degenerate, i.e., one touches another all over the Brillouin zone. (Such a flatband degeneracy was conjectured in frustrated Kagome-3 model but requires unphysical hopping [42, 48]). In the following, we shall refer $\beta_3 = 0$ to *point-touching* (singular) and $\beta_{5,6} = -2$ to *plane-touching* (nonsingular) flat bands. These distinct types of flat bands coexist in the same fractal-like lattice geometry.

Figure 1(a) shows the three fundamental flatband modes (CLS-1, CLS-2 and CLS-3 linked by dark-solid lines), which belong to the two different types of flat bands. Different colored waveguides (sites) show the basic CLSs with different amplitude and phase distributions. The CLS-1 can be selected as a basic CLS for flat band $\beta_3$, the CLS-2 and CLS-3 can be selected as the basic CLSs for flat bands $\beta_{5,6}$. Normally, for a single isolated flat band, it is easy to find a complete set of CLSs because the Bloch eigenstate is always analytic [11, 48]. While, for a flat band with point or plane touching, things get a little more complicated. Obviously, another flatband mode CLS-4 belonging to flat bands $\beta_{5,6}$ can also be found by a counterclockwise $2\pi/3$ rotation from the mode $-$CLS-2 (minus denotes a fully opposite phase, which is also an eigenmode of the considered flat band), just as that the CLS-3 is a clockwise $2\pi/3$ rotation [Fig. 1(d)]. One might intuitively think that one of them must be represented linearly by the other two, but this is not the case, because these three modes are related to each other by rotation rather than translation [48]. The operation between rotation and translation can lead to a conventional state counting mistake in Kagome-3 model with plane-touching flat bands, as has been pointed out recently [42, 48]. One can find more detailed information about the relationship of these three CLSs in our recent work [18]. A linear superposition of them is shown in Fig. 1(e), labeled by the super-CLS. It has the same shape but a different phase distribution compared with the CLS-1 (see Sec. 2 for details).

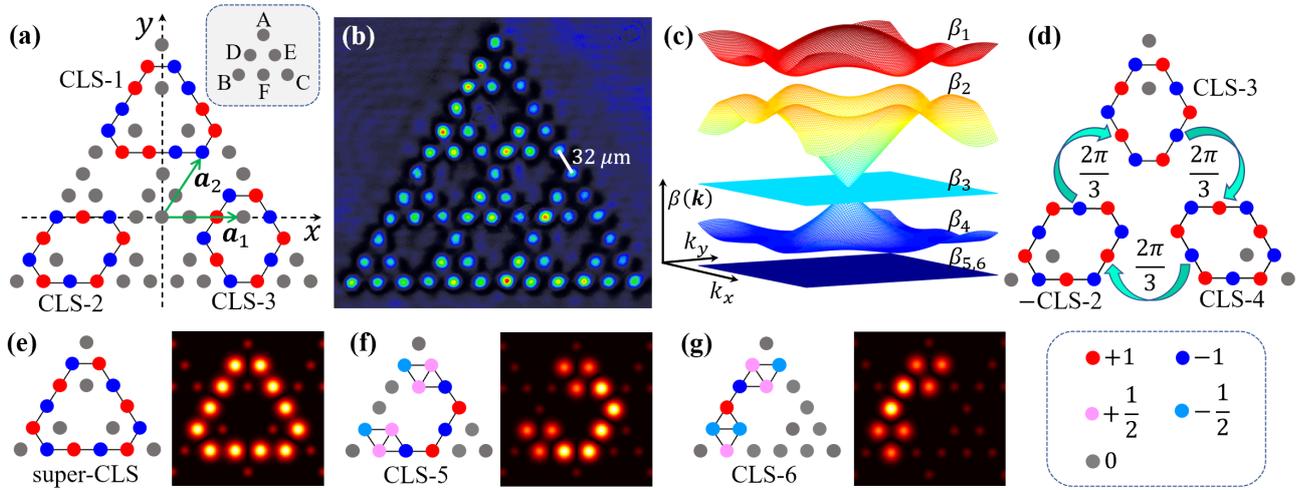

Fig. 1 (a) Schematic diagram of a fractal-like lattice consisting of six sites (marked by A~F in the inset) per unit cell. $\boldsymbol{a}_1$ and $\boldsymbol{a}_2$ marked by green arrows are the primitive vectors in real space. Three examples of the basic CLSs for the two different types of flat bands are shown. (b) An experimentally established laser-writing photonic lattice corresponding to (a). (c) Calculated band structure of an infinite fractal-like lattice under the tight-binding model. From top to bottom, the bands are $\beta_{1\sim6}$, wherein $\beta_3=0$ is a singular flat band with point-touching and $\beta_{5,6}=-2$ are two nonsingular flat bands with plane-touching. (d) Another CLS-4 belongs to the bottom flat band and the relationship among the three CLSs is shown. (e) The super-CLS belongs to the bottom flat band, constructed by a linear superposition of CLS-2, CLS-3, and CLS-4. (f) and (g) The new *armchair-like* CLS-5 (f) and CLS-6 (g) can be obtained by combining rhombic CLS-2 and CLS-3. In (e)-(g), the left panels show their amplitude and phase distributions, the right panels show their intensity patterns. For all figures the different colored sites represent different amplitude and phase distributions. As shown in the dashed box in the right bottom, sites with zero amplitudes are denoted by gray color, normalized amplitude with magnitude 1 are denoted by red and dark blue dots, with magnitude 1/2 by pink and light blue dots, and the sign $+/-$ denotes the opposite phase $0/\pi$.

By a direct linear combination of the rhombic CLS-2 and CLS-3, one can obtain two new *armchair-like* CLSs for the plane-touching flat bands, labeled as CLS-5 and CLS-6, as plotted in Figs. 1(f) and 1(g), respectively. The corresponding unnormalized Bloch wave functions are described by

$$v_{\boldsymbol{k}}^{(5)} \propto \alpha_{\boldsymbol{k}}^{(2)} v_{\boldsymbol{k}}^{(2)} + \alpha_{\boldsymbol{k}}^{(3)} v_{\boldsymbol{k}}^{(3)} = (2e^{ik_3}, 2e^{ik_3}, -2-2e^{ik_2}, -1-2e^{ik_3}-e^{ik_2}, 1+e^{ik_2}, 1+e^{ik_2})^{\mathrm{T}}, \quad (1)$$

$$v_{\bm{k}}^{(6)} \propto \alpha_{\bm{k}}^{(2)} v_{\bm{k}}^{(2)} - \alpha_{\bm{k}}^{(3)} v_{\bm{k}}^{(3)} = \left(2e^{ik_2}, -2, 0, 1-e^{ik_2}, -1-e^{ik_2}, 1+e^{ik_2}\right)^{\mathrm{T}}, \quad (2)$$

where $v_{\bm{k}}^{(2)}$ and $v_{\bm{k}}^{(3)}$ correspond the normalized Bloch wave functions for CLS-2 and CLS-3, $\alpha_{\bm{k}}^{(2)}$ and $\alpha_{\bm{k}}^{(3)}$ are the corresponding normalization factors, respectively (see **SM** for details). Recently, it has been demonstrated that the normalization factor [48, 49]

$$\alpha_{\bm{k}} = \left(\sum_{p=1}^{n} |v_{\bm{k},p}|^2\right)^{1/2} \quad (3)$$

of $v_{\bm{k}}$ can be used as an effective indicator to estimate the completeness of the spanning set of the selected CLS. $p$ denotes the $p$-th component and $n$ the size of the Bloch wave function. The zero point of the $\bm{k}$-dependent $\alpha_{\bm{k}}$ is referred to as the singular momentum [48, 49], also the flatband Bloch wave function's singularity (or flatband singularity for short). The basic verdict is that if $\alpha_{\bm{k}}$ is zero at some possible momenta, the spanning set of the corresponding CLS is incomplete; on the contrary, if $\alpha_{\bm{k}}$ is nonzero at every $\bm{k}$, the corresponding spanning set is complete. Meanwhile, such a zero point is also the very place of the *discontinuity* of the flatband Bloch wave function, which is a direct result of the existence of the singular momentum (more details see **SM**). In what follows, we shall use the terms *discontinuity* and *singularity* interchangeably, as done also in previous publications [48, 49]. Based on these schemes, one can easily diagnose whether a flat band is singular or nonsingular. The singular one means that it is impossible to find a complete set of CLSs to span it, and the nonsingular one means that there exists at least one complete set. More directly, the former usually carries *immovable* singularity in its band structure, while the latter usually without singularity or with *removable* singularities.

One can check that nonzero scalar functions $\alpha_{\bm{k}}^{(5)} = [2(13 + 2\cos k_1 + 7\cos k_2 + 2\cos k_3)]^{1/2}$ and $\alpha_{\bm{k}}^{(6)} = [2(7 + \cos k_2)]^{1/2}$ make $(\alpha_{\bm{k}}^{(5)})^{-1} v_{\bm{k}}^{(5)}$ and $(\alpha_{\bm{k}}^{(6)})^{-1} v_{\bm{k}}^{(6)}$ in the form of a normalized Bloch wave function, respectively. Obviously, the CLS-5 and CLS-6 are also linearly independent with each other. Besides, since the CLS-5 and CLS-6 can be obtained by linear combinations of CLS-2 and CLS-3, there should be many choices with respect to the basic CLSs. Here, however, we only choose these two types of relatively small size, which are obtained by simple and direct summation and subtraction as operations controlled by Eqs. (1) and (2). Of course, for the plane-touching flat bands, one can also choose a CLS-6 and its $2\pi/3$ rotation as the basic CLSs (see **SM**), just as the construction process of the CLS-2 and CLS-3. One can find, in these specific configurations, both the phase and patterns of CLS-5 and CLS-6 [see Figs. 1(f) and 1(g)] are distinctly different with typical CLSs [see Fig. 1(a)]. The reason is that the CLSs in nature are an inverse Fourier transformation of the corresponding Bloch wave functions. Since the CLS-2 and CLS-3 share the same eigenvalue (propagation constant), their superimposed states (CLS-5 and CLS-6) are also eigenstates of the plane-touching flat bands [18, 19]. Thus, for the doubly-degenerate flat bands $\beta_{5,6}$, there exist two different sets of basic CLSs, one set consist of {CLS-2} and {CLS-3}, the other one consist of {CLS-5} and {CLS-6}, where the notation "{·}" indicates $N$ translated copies of the selected basic CLS. When we choose one of the sets, {CLS-2, CLS-3} for example, {CLS-2} corresponds to the flat band $\beta_5$, then {CLS-3} automatically corresponds to flat band $\beta_6$, and vice versa. The same is true when another CLS set {CLS-5, CLS-6} is selected (see **SM for** more details of the CLS-5 and CLS-6).

Let us examine the singularity of the Bloch wave functions of the flat bands $\beta_{5,6}$. Figures 2(a)-2(d) show the normalization factor $\alpha_{\bm{k}}$ as a function of momentum $\bm{k}$, and the zeros indicate the discontinuities of Bloch wave function as stated in Ref. [48]. Obviously, the CLS sets {CLS-2} [Fig. 2(a)] and {CLS-3} [Fig. 2(b)] are incomplete due to the zeros of $\alpha_{\bm{k}}$; while the CLS sets {CLS-5} [Fig. 2(c)] and {CLS-6} [Fig. 2(d)] are complete due to the non-zero values. The incomplete spanning sets usually means there exist some missing states to complete the flatband basis [42, 48]. Previously, it was considered that the NLSs or unconventional line states caused by an *immovable* singularity of the Bloch wave function is a manifestation of real-space topology, such as in Kagome [14, 48], Lieb [10], and super-honeycomb [13] lattices with flatband point touching. In our model with plane touching, the band structure can also present singularities that make corresponding Bloch wave functions vanish. Hence, there should be some missing states: typically, the NLSs. It is worth noting that, the singular momenta of the relevant Bloch wave functions of CLS-2 and CLS-3 are located at different positions [Figs. 2(a) and 2(b)]. This is quite unexpected, and it renders the plane-touching flat bands nonsingular, supporting CLSs (here the constructed CLS-5 and CLS-6) that have no singular momenta [Figs. 2(c) and 2(d)]. The crucial point is that the singularities of the Bloch wave functions of CLS-2 and CLS-3 can be removed by an effective linear combination, i.e., the singularities of the plane-touching flat bands $\beta_{5,6}$ are *removable*. Thus, the resulting new CLSs (CLS-5 and CLS-6) without singularities can span the corresponding nonsingular flat bands.

As discussed above, if one chooses the rhombic CLS-2 and CLS-3 as the basis wave functions to form two independent sets to span the plane-touching flat bands $\beta_{5,6}$ respectively, the introduction of NLSs to complete the flatband basis is required; while, if one chooses the CLS-5 and CLS-6 as the new basis, any introduction of NLSs is redundant. But in any case, the plane-touching flat bands are always nonsingular no matter which basic spanning sets we choose. The former is because the singularities of the corresponding Bloch wave functions are *removable*, and the latter is because the corresponding Bloch wave functions are *continuous* (without singularities) throughout the momentum space. Therefore, for the plane-touching flat bands, the necessity of having NLSs depends on the selection of the spanning sets of the respective flat band. As a product of the Bloch wave function's singularity [48] (or the band-touching from the perspective of state counting [42]), the NLSs serve as a good manifestation of real-space topology, no matter the singularity is *immovable* or *removable*.

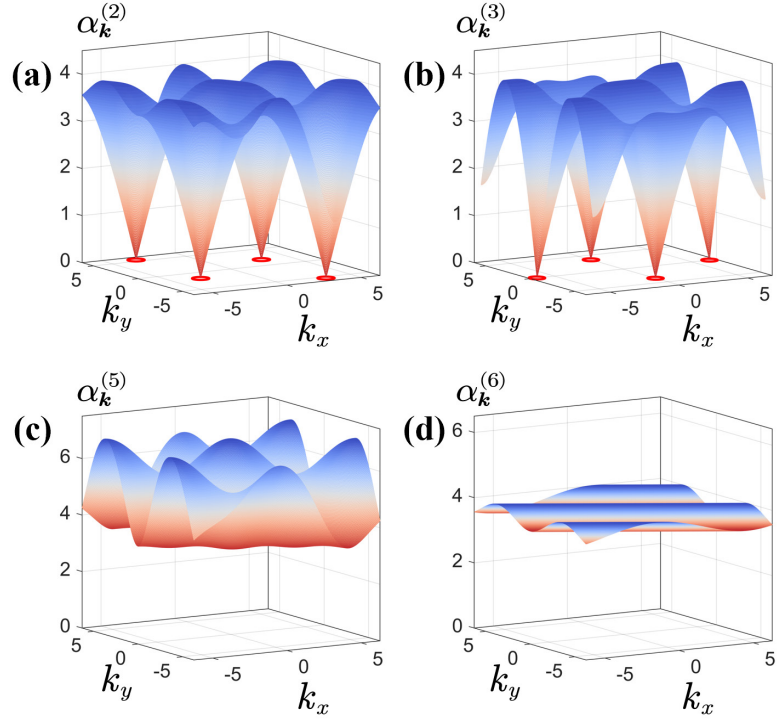

Fig. 2 Illustration of Bloch wave function singularities from calculated normalization factor $\alpha_k$ in momentum space. (a) and (b) correspond to the rhombic CLS-2 and CLS-3, while (c) and (d) correspond to the new *armchair-like* CLS-5 and CLS-6, respectively. The singularities of the relevant Bloch wave functions are manifested by the zero points (red circles) of $\alpha_k$. *Removable* singularities in (a, b) mean incomplete flatband spanning sets {CLS-2} and {CLS-3}, but *no* singularity in (c, d) means complete spanning sets {CLS-5} and {CLS-6}.

*Experimental and numerical results of NLSs from different band-touching*

In this section, we design and establish finite fractal-like photonic lattices to further discuss the NLSs associated with the two types of flat bands, especially the experimental demonstration. Let's first focus on the zero-energy flat band ($\beta_3 = 0$) with singular point band-touching (see **SM**). Since the discontinuity of the Bloch wave function $v_k^{(1)}$ occurs at $\bm{k}=0$ (see **SM** for the expression of $v_k^{(1)}$), $\beta_3$ cannot be completely spanned by the {CLS-1}, and some extended NLSs must be complemented. We know that NLSs usually reside in an ideal torus representing an infinite system with periodic boundary condition [48]. In order to complete the flatband basis in realistic finite-sized lattice system with open boundary condition, the NLSs could be cast into the flatband line states [10, 13, 14, 42, 48], which have been experimentally demonstrated in photonic Lieb [10] and super-honeycomb [13] lattices. Since these implementations are zero-energy flatband eigenstates and present as a "discrete" appearance separated by zero-amplitude lattice sites in real space, they can maintain stable propagation in a finite system with specially truncated boundaries. However, for the photonic fractal-like lattice, one can see that the line-shaped states we calculated for $\beta_3 = 0$ and $\beta_{5,6} = -2$ are both quasi-continuous. The nearest-neighboring lattice sites are not separated by other sites, as displayed in the highlighted areas labelled by line-1, line-2, and line-3 [Figs. 3(a1), 3(b1), and 3(c1)]. This is an indication that these three types of line-shaped states are not ideal flatband eigenstates in a finite system, similar to the previous work in Kagome lattices [14]. Such phenomena usually show the missing couplings of the boundary sites, which can be compensated by introducing dangling lattice sites (see **SM** for detailed information). For the wavy-shaped line-1 state [see Fig. 3(a1)], it is in nature an NLS rooted in the band-touching point under periodic boundary condition, which manifests nontrivially in real space [49, 56] (see **SM**). Therefore, to experimentally observe it we should achieve periodic boundary condition first. By wrapping the finite fractal-like lattice ribbon into a Corbino-shaped geometry [Fig. 3(a2)] and using the site-to-site CW-laser writing technique [10, 14], we generate such a lattice [see Fig. 4(a)] in a biased photorefractive bulk crystal. For the Corbino-shaped geometry, the distances between the sublattices over each ring are equivalent and increase with the ring diameter. The radiuses of the inner and outer rings are 128 μm and 224 μm, respectively. Along the inner ring, the lattice spacing between two adjacent sites is 32 μm. Then, with the strength of this special lattice geometry, a nontrivial loop state located along the toroidal direction, akin to an infinite system with periodic boundary condition in one dimension, can be observed. We launch a wavy-shaped necklace-like pattern [Fig. 4(b)] matching the out-of-phase structure of the NLS-1 [Fig.3 (a2)] as an initial excitation into the lattice and corresponding experimental result is shown in Fig. 4(c1). One can clearly see that the probe beam remains intact after 10-mm propagation through the lattice, as verified also by numerical simulation [Fig. 3(a3)]. For comparison, the input probe beam is reshaped into an in-phase structure. As expected, it is strongly distorted during propagation [Fig. 4(c2)] since such input cannot evolve into the NLS-1. The experimental results are in good agreement with numerical simulations, suggesting the feasibility of our fractal-like photonic systems in the observation of nontrivial NLSs arising from singular point-touching.

Next, we focus on the nonzero-energy flat bands ($\beta_{5,6} = -2$) with plane-touching, which leads to NLS-2 and NLS-3. As discussed above, the two rhombic CLSs [Fig. 1(a)] form incomplete sets for the plane-touching flat bands as manifested by the existence of singularities at $\boldsymbol{k} = (\pi, \pi/\sqrt{3})$, $\boldsymbol{k} = (-\pi, -\pi/\sqrt{3})$, $\boldsymbol{k} = (\pi, -\sqrt{3}\pi)$, and $\boldsymbol{k} = (-\pi, \sqrt{3}\pi)$ in $v_{\boldsymbol{k}}^{(2)}$, and $\boldsymbol{k} = (0, \pm 2\pi/\sqrt{3})$, $\boldsymbol{k} = (\pm 2\pi, 0)$ in $v_{\boldsymbol{k}}^{(3)}$, marked by the red circles [see Figs. 2(a) and 2(b)]. It is interesting that these singularities of the Bloch wave function are not covering the entire Brillouin zone but are located at some discrete momenta. While some of these momenta are equivalent in respective momentum space, the corresponding Bloch wave functions do vanish at their locations. Therefore, additional non-compact (extended) states are needed to complete the incomplete spanning sets [48]. The NLS-2 and NLS-3 shown in Figs. 3(b2) and 3(c2) as the intuitive manifestations of the straight line-2 [Fig. 3(b1)] and wavy-shaped line-3 [Fig. 3(c1)] are independent of the rhombic CLS-2 and CLS-3, respectively. They cannot be constructed by linear superposition of the conventional CLS-2 and CLS-3. Similarly, the two NLSs cannot be disconnected by adding a finite number of CLS-2 and CLS-3 (see **SM**). One obvious result is that the NLS-3 [Fig. 3(c2)] shares the same shape but different phase distributions with the NLS-1 [Fig. 3(a2)], similar to the appearance of the super-CLS [Fig. 1(e)] and CLS-1 [Fig. 1(a)], respectively. Also, we have experimentally demonstrated the NLS-2 and NLS-3 for the plane-touching flat bands and typical experimental results after 10-mm propagation through the lattice are shown in Figs. 4(d1) and 4(e1), respectively. The output patterns show that the probe beams are well localized under out-of-phase excitations but exhibit discrete diffraction for the in-phase cases [Figs. 4(d2) and 4(e2)]. The experimental results are in good agreement with numerical simulations as shown in Figs. 3(b3) and 3(c3). In a word, all the three NLSs are compact localized along the radial direction while extended along the azimuthal direction, they are truly noncontractible and have identical observable properties in real space.

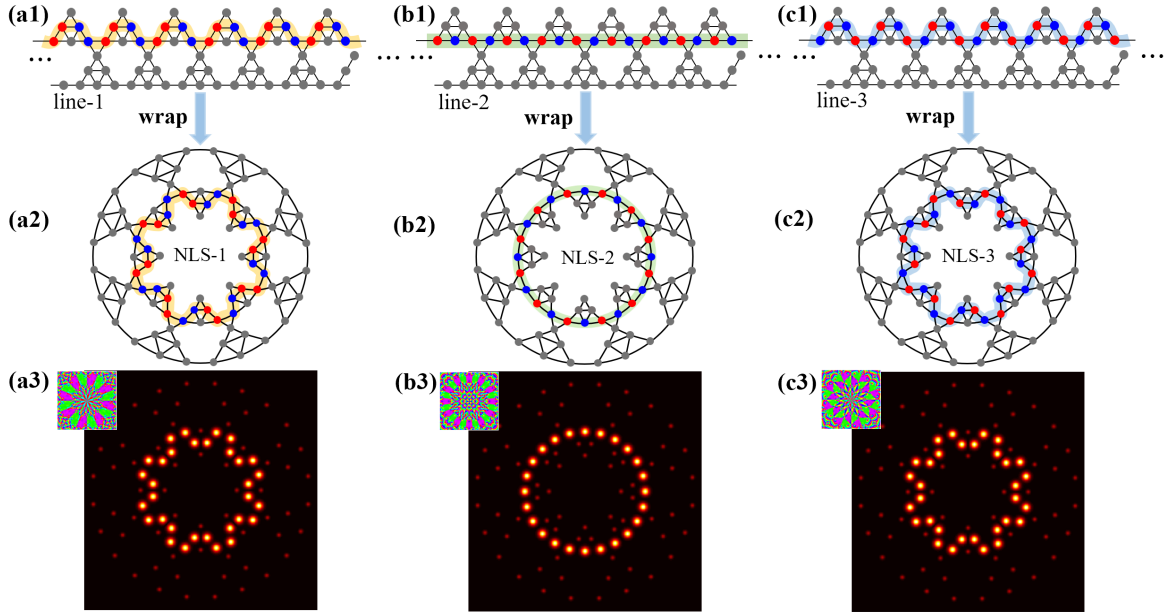

Fig. 3 (a1) A wavy-shaped line state (marked by the orange line) supported by a fractal-like lattice ribbon. (a2) A Corbino-shaped fractal-like lattice by warping the ribbon into a ring, where the orange loop illustrates an NLS (NLS-1 at $\beta_3 = 0$). (a3) The NLS-1 obtained in simulations at a propagation distance of 10 mm under out-of-phase condition. The inset on the top left corner shows the corresponding phase distribution. (b1)-(b3) and (c1)-(c3) have the same layout as (a1)-(a3), except that they belong to the nonsingular flat band at $\beta_{5,6} = -2$ with plane-touching. In all the figures, the representation of the colored sites is the same as in Fig. 1.

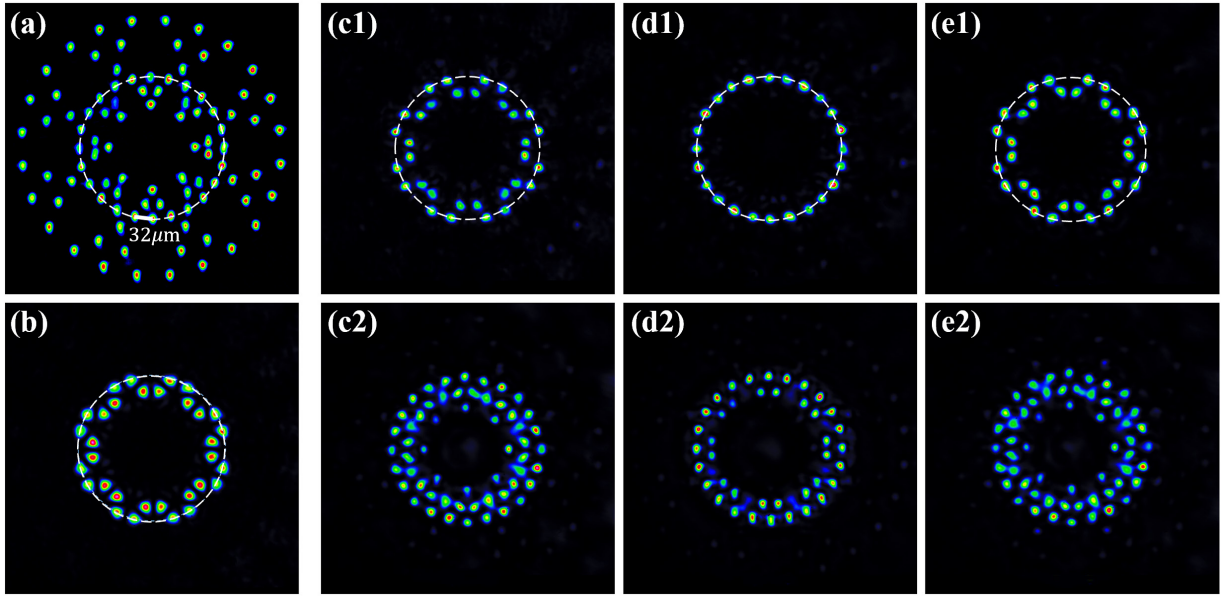

Fig. 4 Experimental results of three distinct types of NLSs in Corbino-shaped fractal-like photonic lattices corresponding to those in Fig. 3. (a) Experimentally established lattice with the CW-laser-writing method. The dashed ring marks the excitation location of the inner ring. (b) The shape of the input probe beam for (c1) and (e1), for which the phase relation has been shown in Figs. 3(a2) and 3(c2), respectively. (c1), (d1), and (e1) are the output patterns after 10-mm propagation through the lattices under out-of-phase excitations. (c2), (d2), and (e2) are the corresponding output patterns under in-phase excitation for direct comparison.

Until now, we have observed the two types of distinct NLSs with different origins in Corbino-shaped fractal-like photonic lattices. One type (NLS-1) originates from the singular point band-touching, while the other type (NLS-2 and NLS-3) originates from the nonsingular plane band-touching. The fundamental and direct origin of the NLSs is the existence of Bloch wave function's singularities, both *removable* and *immovable* [48], which makes the relevant CLSs cannot form a complete set spanning the flat band. A rigorously theoretical analysis for the existence of the NLSs in the Corbino-geometry fractal-like lattices from the perspective of state counting can be found in **SM**. Besides, one may find that the NLS-2 and NLS-3 are noncontractible for the rhombic states (CLS-2 and CLS-3) but contractible for the new *armchair-like* states (CLS-5 and CLS-6) (more details, please see **SM**). This does not mean the NLS-2 and NLS-3 presented here are not truly noncontractible. Instead, it exactly solidifies the proposal that the necessity of the existence of independent NLSs in the plane-touching flat band hinges on the selection of the basic CLSs sets. There is an undeniable fact that although NLS-2 and NLS-3 are not independent of the new CLS-5 and CLS-6, they are indeed independent of the rhombic CLS-2 and CLS-3. One can get a better understanding for this point in the following section.

*Robust boundary modes as indirect manifestations of NLSs*

In flatband lattices under open boundary conditions, the robust boundary modes (RBMs), as an emerging nontrivial localized state, represent NLSs indirectly and have been experimentally realized in Kagome photonic lattices with a singular point-touching [14, 48]. They are conceived from a torus geometry (infinite system with periodic boundary condition) by cutting edges along the poloidal and toroidal directions (turn into a finite system with open boundary condition) and therefore bridge the gap between infinite and finite lattice systems [48]. The meaning of "robust" in RBMs has at least two aspects: First, the RBMs are robust to random disorders of amplitude or phase of a finite number of sublattices [14]. This robustness is largely dependent on the destructive-interference structure obtained by summing a macroscopic number of CLSs [48, 49]. In fact, even a single CLS can also exhibit such robustness [8]. Second, the RBMs were previously thought to be derived from the singular band touching and cannot be cut off by adding a finite number of CLSs, i.e., the RBMs have the same origin as the NLSs [48, 49]. Such RBMs are well studied in flatband systems with singular point-touching, while that with nonsingular plane-touching have remained unexplored, and will be unveiled here.

We first construct three types of RBMs in finite fractal-like photonic lattices with different open boundaries, as sketched in Figs. 5(a)-5(c), RBM-1 with four wavy-line boundaries, RBM-2 with four straight-line boundaries, and RBM-3 with two wavy-line and two straight-line boundaries. All RBMs discussed here are flatband eigenmodes, and they exhibit identical observable properties in real space. However, their origins are different: The RBM-1 is supported by the *immovable* singularity of the point-touching flat band, while the other two are supported by the *removable* singularities of the plane-touching flat bands. They are constructed by collecting corresponding basic CLSs via a direct or staggered sum. Each of these three boundary modes consists of nine CLSs identified by nine triangular plaquettes, the number of the total lattice sites they occupied are 44, 34, and 38, respectively. It is important to note that while these RBMs are eigenstates of the flat bands and related to the singularities of the Bloch wave function, they are not independent of the corresponding component CLSs. More directly, they are more like the super-CLS shown in Fig. 1(e), which is also constructed by multiple basic CLSs. Here, we further realize that the super-CLS is essentially an RBM on a small scale consisting of only three unit cells. As depicted in Figs. 5(d1)-5(d3), the RBM-2 can merely be deformed after adding a CLS-2, whose $N$ translated copies form an incomplete spanning set. On the contrary, after adding a linear superposition of two CLS-6s ($N$ translated copies of CLS-6 form a complete spanning set), one can see that the RBM-2 is easily cut off [Figs. 5(e1)-5(e3)]. Does this mean that the RBM is not "robust"? Of course not! This only shows that the RBM-2 comes from the *removable* singularity of the Bloch wave function of the CLS-2, which can naturally be cut by the CLS-6 that can generate complete spanning sets. Moreover, it can be checked that the broken RBM in

Fig. 5(e3) is still localized well during propagation. From this point of view, the RBM in a finite system shows more the properties of the CLSs that makeup it. But in any case, RBMs can always be used as an indirect manifestation of NLSs.

In short, we demonstrate that although the RBM or NLS is a crucial signature for identifying the singular point-touching of the flat band [14, 48, 49], it is not peculiar to the singular flat band. The nonsingular plane-touching flat band also supports such RBMs and NLSs even with more patterns, as shown in Figs. 5(b,c) and 4(d1,e1). From this point of view, a flatband system with plane-touching has more significant advantages in realizing large-scale image transmission when compared with the point-touching flatband system [7, 10]. Different from the general topological edge modes supported by the momentum-space topology, the RBMs and NLSs derived from the point- and plane-touching flat bands are supported by the nontrivial real-space topology caused by special lattice geometries. Both of the point- and plane-touching of the flat bands support a novel bulk-boundary correspondence such that the presence of the RBMs and NLSs is guaranteed by the singularities of the Bloch wave functions of the bulk CLSs.

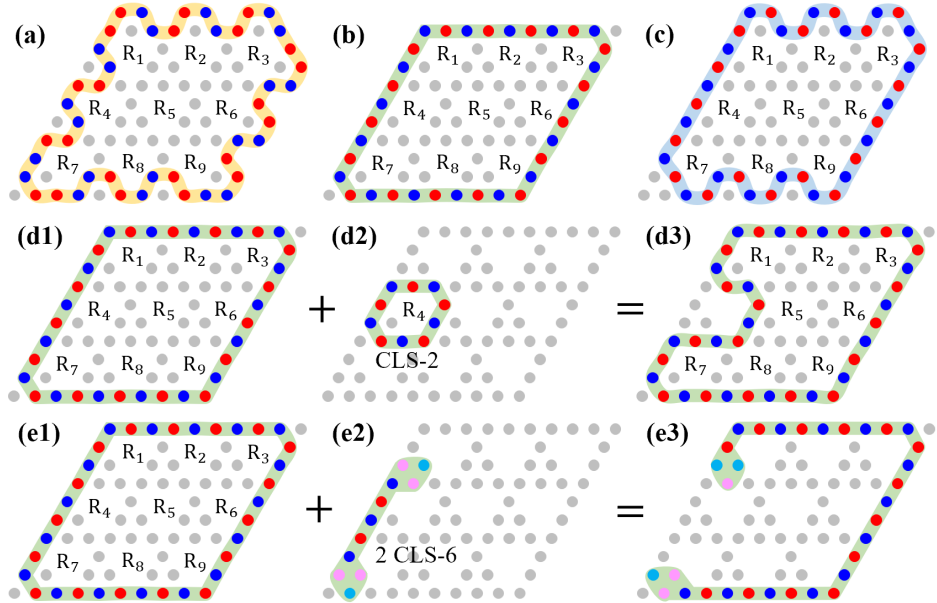

Fig. 5 Three configurations of RBMs with different open boundaries. Each of them can be obtained by adding all the CLSs centered at all the unit cells (from $R_1$ to $R_9$). (a) RBM-1 with four wavy-line boundaries. (b) RBM-2 with four straight-line boundaries. (c) RBM-3 with two wavy-line and two straight-line boundaries. Note that the RBM-1 also supports the alternating phase distributions as RBM-2 and RBM-3. (d1)-(d3) show how an RBM-2 [from (b)] is deformed (d3) by adding a CLS-2 in (d2). (e1)-(e3) show how the same RBM-2 is "cut" and disconnected (e3) by adding two CLS-6s in (e2). In all the figures, the representation of the colored sites is the same as in Fig. 1.

Finally, we would like to point out the significance of aforementioned basis selection considering both conventional flatband state counting [42] and emerging Bloch wave function's singularity [48]. As we know, the presence or absence of flatband touching can be understood by a careful counting of linearly independent CLSs. Crucial to this counting is the presence of NLSs, a direct manifestation of real-space topology [42]. The isolated plane-touching flat band possesses two features: One originates from the internal-band double degeneracy, the other originates from the external gapped (without band touching) characteristic with other bands. Previous discussions on an isolated plane-touching flat band were based on an unrealistic Kagome-3 lattice model, neglecting either the external properties [42] or the internal properties [48]. Here, by exploring a similar case but based on the realistic fractal-like lattice, we argue that the internal property is exactly reflected in the presence of the additional NLS-2 and NLS-3 when we choose the CLS-2 and CLS-3 as a basis; the external (isolated) property is exactly reflected in the absence of NLSs when we choose the CLS-5 and CLS-6 as a basis (see **SM**). Interestingly, the results from the perspective of Bloch wave function's singularity are in perfect agreement with those from the state counting. This further illustrates that the existence of the *removable* singularities is a unique feature of the plane-touching flat bands.

Before closing, we would like to emphasize again the difference of this work with respect to those reported in previous publications [18, 19]. First, our current work focuses on two types of NLSs and their origins, while those previous work mainly concentrated on the basic CLSs. Second, NLSs are traditionally believed to come from the singular point-touching between a flat band and another dispersive band, as pictured also in Ref. [19], but we show here that nonsingular plane-touching of the flat bands can also give rise to NLSs. Therefore, our current work on NLSs and their direct and indirect manifestations differ from those previous work. Only with a full understanding of the singularity of the Bloch wave function (or the completeness of the CLS) can we gain a deeper understanding of the origin of the NLSs. We therefore hope that the intriguing questions we addressed here will stimulate further interest in exploring flatband physics beyond photonics.

## 3. Conclusion

In conclusion, we have demonstrated a few distinct types of topological NLSs and RBMs in fractal-like photonic lattices with co-existing point (singular) and plane (nonsingular) band touching. The results of this work will not only deepen our understanding of the origin of the nontrivial flatband states and real-space topology but also provide insight into complex and interesting many-body localized states [57-59]. The singular point-touching presented here consists of a pseudospin-1 Dirac-like cone, where numerous intriguing phenomena related to momentum-space and real-space topology can be investigated [35, 54, 60]. Meanwhile, the isolated flat bands with plane-touching will call for research on fundamental issues such as anomalous Landau level spreading [30, 31]. It is noted that the Sierpinski fractal-like lattice inherently contains frustrated Kagome-type geometry, which is already well-established and has been realized in various materials [24, 42, 49, 61]. This opens the way for the extensive realization of fractal-like lattices in other physical settings by similar schemes. Furthermore, our work provides a reference for the study of complex strongly-correlated phenomena in multi-flatband systems involving isolated flatbands of plane-touching degeneracy. We envisage that the concepts and results presented in this work can be applied to other flatband systems beyond photonics, including for example electronics, acoustics, microwaves, electric circuits, and cold atoms.

**Acknowledgment**

This work was supported by National Key R&D Program of China (2017YFA0303800); the National Natural Science Foundation of China (12134006 and 11922408); the Natural Science Foundation of Tianjin (21JCYBJC00060); the Natural Science Foundation of Tianjin for Distinguished Young Scientists (Grant No. 21JCJQJC00050), PCSIRT (IRT_13R29), 111 Project (No. B07013) in China.

**Conflict of Interest**

The authors declare no conflict of interest.

**Keywords**

nontrivial loop states, fractal-like lattice, singularity, singular and nonsingular flat bands